\documentclass[%
 reprint,
superscriptaddress,
preprintnumbers,
 amsmath,amssymb,
 aps,
prb,
floatfix,
]{revtex4-2}

\usepackage{graphicx}
\usepackage{dcolumn}
\usepackage{bm}
\usepackage{xcolor}
\usepackage{hyperref}
\usepackage[mathlines]{lineno}

\begin{document}


\title{Light dark bosons in the JUNO-TAO neutrino detector}

\author{Mikhail Smirnov}
  \email{gear8mike@gmail.com}
 \affiliation{School of physics, Sun Yat-Sen University, Guangzhou 510275, China}
 \affiliation{Friedrich-Alexander-Universit\"{a}t Erlangen-N\"{u}rnberg, Erlangen Centre for Astroparticle Physics, Erlangen 91058, Germany}

\author{Guang Yang}
\email{guang.yang.1@stonybrook.edu}
\affiliation{Department of Physics and Astronomy, State University of New York at Stony Brook, \\Stony Brook, New York 11794, USA}
 
\author{Jiajun Liao}
\email{liaojiajun@mail.sysu.edu.cn}
\affiliation{School of physics, Sun Yat-Sen University, Guangzhou 510275, China}

\author{Zhuojun Hu}
\email{huzhj3@mail2.sysu.edu.cn}
\affiliation{School of physics, Sun Yat-Sen University, Guangzhou 510275, China}

\author{Jiajie Ling}
\email{lingjj5@mail.sysu.edu.cn}
\affiliation{School of physics, Sun Yat-Sen University, Guangzhou 510275, China}

\date{\today}

\begin{abstract}

This work presents a sensitivity study of a reactor liquid scintillator detector to three kinds of dark bosons with masses below 1 MeV, such as dark photons, axion-like particles and light scalar bosons. The JUNO-TAO detector with Taishan nuclear reactor is taken as a reference. With proposed 180 days data taking, \textcolor{black}{the best sensitivity result at level of $\sim10^{-5}$ 95\%C.L. is achieved for dark photons with} an optimized signal to background ratio for the electron coupling constant $\it{g_X} $ through inverse Compton-like scattering. 
\textcolor{black}{
Similar calculations are completed for axion-like particles and scalar bosons.
}
The background systematic uncertainty presents as the main limiting factor for the further sensitivity improvement. 
Several remarks are made to the controversial analysis for the NEOS experiment.
Additionally the differential and the inverse differential cross sections have been derived for all three boson types and their interactions with electrons in liquid scintillator.
\end{abstract}

\maketitle

\section{Introduction}

Neutrino experiments have provided us a lot of information about the Standard Model (SM) and the physics beyond it.
Reactor experiments have become an important part of neutrino experiments recently since nuclear reactors can provide us an intensive antineutrino flux.
Experiments such as KamLAND~\cite{kamland}, Daya Bay~\cite{dyb}, RENO~\cite{reno} and Double Chooz~\cite{Abe:2011fz} have been successful in measuring neutrino oscillations and all of them use liquid scintillator (LSc) as a target material. 
On the other hand, experiments such as NEOS~\cite{neos}, DANSS~\cite{Alekseev:2018efk}, STEREO~\cite{stereo}, PROSPECT~\cite{prospect}, Neutrino-4~\cite{n4} and SoLid~\cite{solid} aim to search for sterile neutrinos.
These experiments have a short baseline (usually a few tens of meters) to the neutrino source in order to measure the distortion of the antineutrino energy spectrum due to the possible existence of sterile neutrinos.
Such experiments can be used to search for popular candidates of dark matter particles.
In this paper, instead of focusing on the dark matter review and some theoretical aspects, we intend to present the requirements  for the LSc-based reactor neutrino detectors to search for certain dark matter candidates. 
Some discussions about previous research in this area will be presented as well.
As an example, we present the sensitivity results for the under-construction TAO detector (Taishan Antineutrino Observatory)~\cite{tao}, which serves as JUNO's near detector~\cite{juno}.

Inclusion of dark matter is a necessity from many observed phenomena, especially in cosmology, which can not be fully explained in the framework of SM.
Theories that modify  gravity usually gain success in local regions but fail at  larger scales, thus dark matter is in the spotlight of the entire physics community.
The number of possible candidates and theories is expanding.
Recently one of the main candidates for dark matter, WIMPs (weakly interacting massive particles), was ruled out by the XENON1T experiment at the level down to a cross section of $\sigma\approx4.1\cdot10^{-47}$ $\rm cm^2$ at 90\%C.L.~\cite{Aprile:2018dbl} and incoming projects such as PandaX aims for a tighter constraint~\cite{Meng:2021}.
A search for alternative candidates is critical.
As is known, there is an absence of large accumulations of dark matter particles (GeV scale) in the solar system.
The main clumps of dark matter are located much further in the Milky Way galaxy.
It is likely that these particle can interact with ordinary matter only through the gravitational forces.  
All expected properties of dark matter particles lead to an extremely difficult direct observation. 
At the same time dark matter particles may have self-interactions with each other.
Such interactions can be realized through the exchange of intermediate bosons by analogy with ordinary interactions between SM particles.
Mass range of the dark bosons is not restricted, but if their mass is  below than $2m_e$ they are not allowed to decay to electron-positron pair and thus they could be stable.
Interaction between  SM particles and dark bosons opens a so-called ``portal'' through which the dark matter particles can be observed.
There are various portals and the relevant ones are described below.
\begin{itemize}
\item {\it Vector portal} is represented by a hidden photon, also named as a dark photon (DP)~\cite{Holdom:1985ag}. 
The DP can be kinetically mixed with the electromagnetic photon.
The strength of the kinetic mixing is determined by a coupling constant $g_e$. \textcolor{black}{In general, the DP could be a very weakly coupled light gauge boson~\cite{Fayet:1980ad,Fayet:2007ua}.}
Moreover, DP should have mass by analogy with a Higgs-like mechanism.
\item {\it Axion portal} can include pseudoscalar axions or axion-like particles (ALP).
The axion was suggested to solve the strong CP problem~\cite{Peccei:1977hh} and it was later considered as a candidate of the dark matter. 
\item {\it Scalar portal} can be attributed to the light Higgs bosons or scalar axions (or scalar boson, SB)~\cite{Filippi:2020kii}.
\end{itemize}
All these three types of bosons can be massive and produced in Compton-like scattering. 
Other possible production and detection channels like the one through Primakoff effect~\cite{Pirmakoff} for ALPs are not considered in this research for simplicity. We only consider dark bosons interacting with electrons, thus they can be produced by Compton-like scattering and detected by its inverse reaction.

In addition to Compton-like scattering, various observables have been used to constrain the strength of interaction between the low-mass boson and the electron. 
Dark matter experiments have been searching for ALPs via electronic recoil~\cite{Armengaud:2018cuy,Aprile:2020tmw}. 
Searches of photon to DP and then DP to photon conversions such as CAST~\cite{Redondo:2008aa}, CROWS~\cite{Betz:2013dza}, Light Shinning through a Wall (LSW)~\cite{Ehret:2010mh} and SHIPS~\cite{Schwarz:2015lqa} set bounds on the light ALPs and DPs parameters. 
Direct dark matter detection (XENON10~\cite{An:2013yua}) result can also be translated into bounds on DPs. 
The same is true for atomic experiments~\cite{Jaeckel:2010xx} searching for modifications of the Coulomb's law due to DPs. 
In astrophysics and cosmology, the absence of deviations on the black body spectrum in the Cosmic Microwave Background (CMB)~\cite{Mirizzi:2009iz,McDermott:2019lch,Caputo:2020bdy,Garcia:2020qrp}, decays of relic dark matter~\cite{Redondo:2008ec} and anomalous energy transport in stars~\cite{Redondo:2013lna} provide powerful constraints on DPs. 
For light SBs, precision atomic \textcolor{black}{and optical} experiments~\cite{Antypas:2019qji,Arvanitaki:2016fyj, Stadnik:2014tta, Grote:2019uvn, Vermeulen:2021epa, Aiello:2021wlp} have been utilized to search for effects on the fundamental constants induced by scalar dark matter.

In the following sections, a description of the theoretical scattering processes of hypothetical particles is presented, followed by an introduction to the experimental setup, an analysis framework description and results with discussions. 

\section{Compton-like scattering and inverse scattering on electrons}

Compton-like scattering and its inverse reaction are key processes in the analysis. 
The total Compton-like cross section for all three types of hypothetical particles can be found in~\cite{Gondolo:2008dd, Dent:2019ueq, AristizabalSierra:2020rom}.
On the other hand, in order to complete a detailed calculation of the event rate the differential interaction cross sections are necessary. 
Thus authors of this paper have derived differential cross sections for the bosons mentioned in the previous section.

Compton-like scattering is a process that an incident gamma photon interacts with a bound electron in a target atom and then it converts to an outgoing massive boson.
In the case of a vector interactions, the Lagrangian involving dark bosons $X$ can be written as:
\begin{equation}
\label{eq_1}
\mathcal{L} \supset  -\frac{1}{4} X_{\mu\nu}X^{\mu\nu}+\frac{1}{2}m_X^2 X^2- g_X \bar{e} \gamma^\mu e X_\mu,
\end{equation}
where $X_{\mu\nu}\equiv\partial_\mu X_\nu-\partial_\nu X_\mu$, $g_X$ can be related to the kinetic mixing parameter $\varepsilon$ by $g_X\equiv-\varepsilon e$, $m_X$ is the dark boson mass.
The gamma energy threshold of the reaction $\gamma + e^-\rightarrow X + e^-$ is:
\begin{equation}
\label{eq_2}
 E_{\gamma}^{th}=m_X + \frac{m_X^2}{2m_e}.
\end{equation}
It should be noted if the energy of the incident photon is much lager than~\eqref{eq_2}, then the cross section does not depend on $m_X$. 
The differential cross section for Compton-like scattering on a bound electron can be expressed as:
\begin{widetext}
\begin{equation}
\begin{split}
\label{eq_3}
\frac{d\sigma_{\gamma e^-\rightarrow X e^-}}{dx}=\frac{Z\alpha g_X^2}{2 (s-m_e^2)^3(1-x)}\bigg[-2 m_e^2s(x^2+2)+&
2m_X^4-2m_X^2 s x +s^2(x^2-2x+2)\\
&-\frac{m_e^4(x^3-3 x^2-2)+2 m_e^2m_X^2(x-2)}{1-x}\bigg],
\end{split}
\end{equation}
\end{widetext}
where $s$ is the Mandelstam variable ($s=m_e^2+2E_\gamma m_e$), $Z$ is the atomic number of target material, $\alpha$ is the fine structure constant and $x$ is the fractional light cone momentum~\cite{Dent:2019ueq}.
In the laboratory frame it can be written as $x=1-E_X/E_\gamma+m_X^2/(2 E_\gamma m_e)$, where 
$E_X$ is the total energy of the dark boson (DP in this case).
The differential cross section with respect to the electron recoil energy $T$ or the $E_X$ can be directly obtained from~\eqref{eq_3}:
$$\frac{d\sigma_{\gamma e^-\rightarrow X e^-}}{dT}=\frac{d\sigma_{\gamma e^-\rightarrow X e^-}}{dE_X}=\frac{1}{E_\gamma} \frac{d\sigma_{\gamma e^-\rightarrow X e^-}}{dx},$$
while the total cross section of this reaction is:
\begin{equation}
\label{eq_4}
\sigma_{\gamma e^-\rightarrow X e^-}=\frac{Z\alpha g_X^2}{8s^2(s-m_e^2)^3}\left[A+B\cdot\log C\right],
\end{equation}
where $A, B, C$ are shown in the~\hyperlink{appendix1}{Appendix} due to complexity. The equation~\eqref{eq_4} is in good agreement with the result from~\cite{Gondolo:2008dd}. 

Inverse Compton-like scattering can be derived in a similar manner.
This is a process where the incident dark boson interacts with an atomic electron and is converted into an ordinary photon.
The exact equation of the differential cross section for $X+e^-\rightarrow \gamma+e^-$ together with other two cases, ALPs and SBs, are also detailed in the~\hyperlink{appendix1}{Appendix}.

\section{Experimental setup}
The discovery of new physics requires sufficient validations across different experiments. 
In order to understand the capabilities of comparable experiments better, correct sensitivity projections are critical.
One of the purposes of this paper is to provide a direct comparison with the controversial result of searching for DPs with the NEOS detector~\cite{park}.

\subsection{Nuclear reactor as a source of dark bosons}

The dark bosons are assumed to be generated through the photon electron scattering.
Nuclear reactor is a source of high-intensity MeV photons and it has a large number of electrons.
Therefore, we propose using the nuclear reactor located in Taishan nuclear power plant (NPP) in Guangdong, China as our source of the dark bosons.
It is one of two reactors at the distance of 53 km to the JUNO detector~\cite{juno}.
The Taishan NPP has four Evolutionary Power Reactor (EPR) cores with 4.6 GW thermal power of each. 
At the moment, only two cores are in operation.
The gamma flux measurement has not been fully carried out by Taishan NPP. 
Hence, due to the similarities, the gamma flux is taken from the previous researches such as the FRJ-1 reactor core~\cite{Bechteler} and the TRIGA nuclear reactor core~\cite{Agnolet:2016zir}.
The flux of photons from the reactor with power $P$ has an exponential shape and can be simply presented as~\cite{park}:
\begin{equation}
\label{eq_5} 
\frac{dN_\gamma}{dE_\gamma}=0.58\cdot10^{18}\cdot P(\rm{MW})\cdot\exp[-1.1\cdot E_\gamma(\rm{MeV})].
\end{equation}
Energy window for gammas is from 0.8 to 10 MeV, which will be explained in more details below. 

\subsection{Antineutrino detector}

As an example, we consider TAO, which will serve as the JUNO's reference detector.
It has a distance of 30 m to the reactor core of the Taishan NPP .
The detector has a spherical shape with two nested spheres.  
The inner acrylic vessel is filled with gadolinium doped LAB scintillator~\cite{Beriguete:2014gua} and is surrounded by a copper shell with a diameter of 1.8 m. 
The total mass of the scintillator target is 2.8 tons.
The light yield is around 4500 photoelectrons per MeV given the excellent coverage of the silicon photomultipliers (SiPM).
In order to reduce the dark noise of SiPMs, the central detector operates at $-50^{\rm o}$C.
The unprecedented energy resolution of 1.5\% at 1 MeV and 5 cm position resolution enable precise measurements of the reactor antineutrino spectrum and other quantities in the TAO project~\cite{tao}.
In the detector hall, muon and cosmogenic neutron rates are measured to be 1/3 of those on the ground.

\section{Analysis framework}

Having presented the particle source and detector in the previous section, we  summarize the parameters of the experimental setup in Table~\ref{tab_1} and then move on to the dark boson production and detection rates in this section. 
The detectable single events induced by dark bosons are our signal.
The sensitivity of the detector to those particles depends on the signal to background ratio.
\begin{table}[h]
\caption{\label{tab_1}
A table with input parameters of the proposed experiment.
}
\begin{ruledtabular}
\begin{tabular}{cccc}
\textrm{Mass, ton}&
\textrm{Distance, m}&
\textrm{Power, GW}&
\textrm{Exposure, days}\\
1.9 & 30 & 4.6 & 180\\
\end{tabular}
\end{ruledtabular}
\end{table}

\subsubsection*{Production rate}
\begin{figure}[ht]
\centering
\includegraphics[scale=0.455]{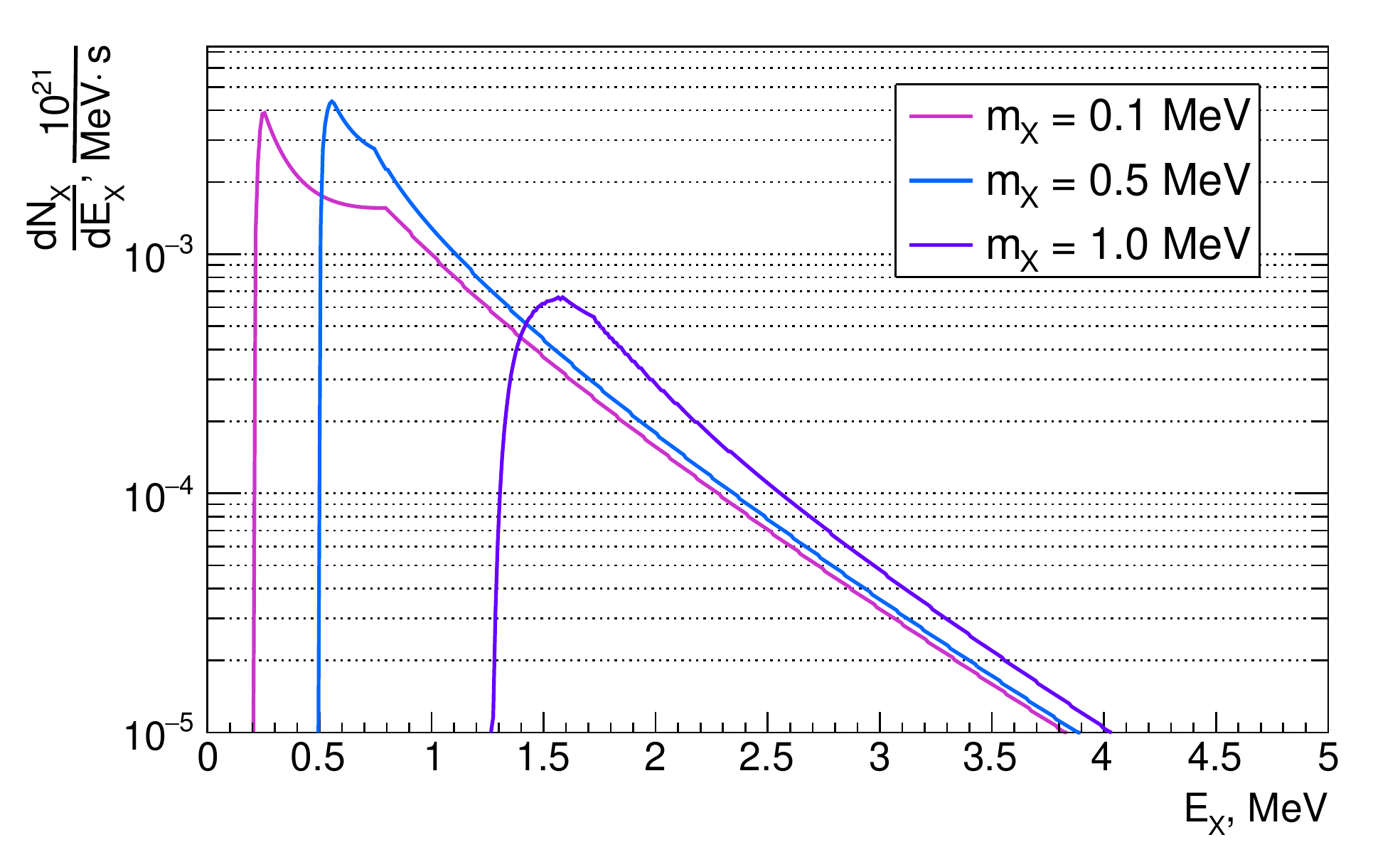}
\caption{ \label{fig_1}The production rate of DPs per second for 1 GW reactor power. Three different masses of DPs are assumed. The coupling constant $g_X$ equals to one. }
\end{figure}

A gamma photon, which was produced inside the reactor with energy at the chosen window is considered.
The photon may interact through Compton-like scattering with an electron in the reactor core material.
In this process a dark boson with mass $m_X$  and total energy $E_X$ is generated in the final state.
The differential rate for $X$ production is given by  equation:
\begin{equation}
\label{eq_6}
\frac{dN_X}{dE_X}=\int \frac{1}{\sigma_{tot}+\sigma_{\gamma e^-\rightarrow X e^-}}\cdot \frac{d\sigma_{\gamma e^-\rightarrow X e^-}}{dE_X}\cdot \frac{dN_\gamma}{dE_\gamma}dE_\gamma,
\end{equation}
where $\sigma_{tot}$ is the total cross section of the interaction between photons and the reactor core material. 
In our study we assume that the reactor core consists of pure Thorium ($Z=90$)~\cite{Dent:2019ueq}.
Compton scattering dominates in the energy range between 0.8 to 10 MeV.
The $\sigma_{tot}$ is calculated numerically~\cite{data_compton} and the other terms in equation~\eqref{eq_6} are obtained based on the aforementioned information.
In addition, the following kinematic condition for $E_\gamma$ in equation~\eqref{eq_6} should be fulfilled:
$$E_\gamma(E_X)=\frac{-0.5m_X^2+E_Xm_e}{m_e-E_X+\sqrt{E_X^2-m_A^2}\cdot\cos\theta},$$
where $\theta$ is the angle is between the incident photon and the outgoing dark boson.
The production rate of DPs is shown in Figure~\ref{fig_1}.
Distortions in the curves for dark bosons with energies less than 0.8 MeV are due to the lower limit of 0.8 MeV for $E_\gamma$ while the exponential slope of equation~\eqref{eq_5} favors low energy photons. 

\subsubsection*{Detection rate}

\begin{figure}[ht]
\centering
\includegraphics[scale=0.455]{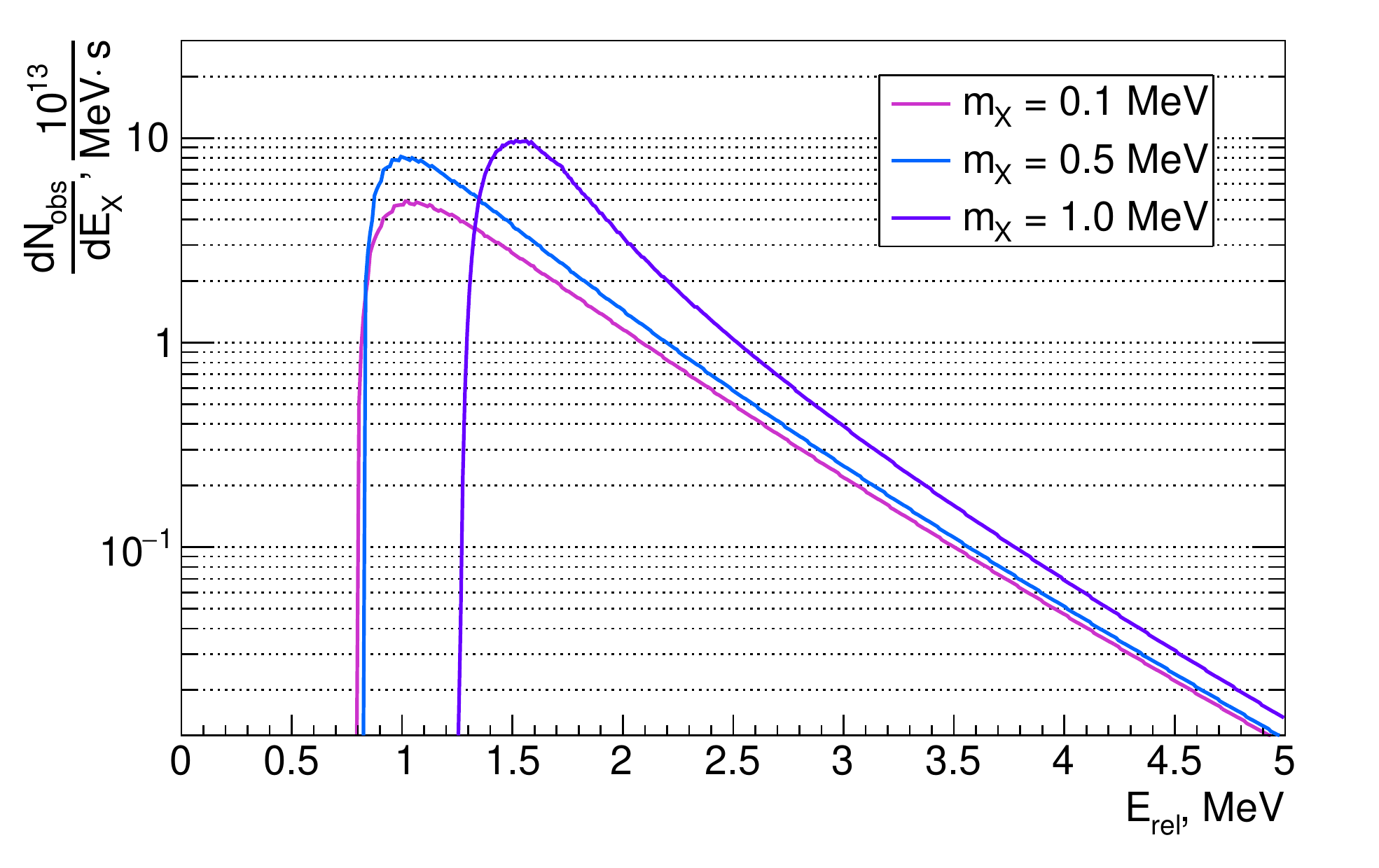}
\caption{ \label{fig_2}The detection rate of DPs per second for a 1 GW reactor. Three different masses of DPs are presented. Coupling constant $g_X$ equals to one. Fiducial mass of LSc is 1 ton. Distance $R$ is 30 m.}
\end{figure}

Dark bosons may decay in-flight after escaping from the reactor core via three possible decay modes: photons, neutrinos or electron-positron pairs.
The decay channels of the dark boson decaying to two or three photons are highly suppressed given the short travel distance between the reactor and the detector~\cite{park}.
In addition, since the $m_X$ is less than $2m_e$ the decay mode of an electron-positron pair is prohibited. Lastly, the decay mode of two neutrinos is not considered in this study.
For simplicity, oscillations between dark bosons and ordinary photons~\cite{Danilov:2018bks} are also omitted in this paper. 
Without any inclusion of the dark boson decays, the detection rate can be written as the following:
\begin{equation}
\label{eq_7}
\frac{dN_{\rm obs}}{dE_X}=\frac{N_eT}{4\pi R^2}\cdot\int\frac{d\sigma_{X e^-\rightarrow \gamma e^-}}{dE_{\gamma '}}\cdot \frac{dN_X}{dE_X}dE_{\gamma '},
\end{equation}
where $N_{\rm obs}$ is the number of observed single events in the detector, $N_e$ is the electron density (for LAB $3.51\cdot10^{29}$ per 1 ton~\cite{Wurm}), $T$ is the exposure time, $R$ is the distance between the reactor core center and the detector, $d\sigma_{X e^-\rightarrow \gamma e^-}/dE_{\gamma '}$ is the inverse Compton-like differential cross section and $E_{\gamma '}$ is energy of the outgoing gamma photon.
The kinematic condition for equation~\eqref{eq_7} is as follows:
$$E_{\gamma '}(E_X)=\frac{0.5m_X^2+E_Xm_e}{m_e+E_X-\sqrt{E_X^2-m_A^2}\cdot\cos\phi},$$
where $\phi$ is the angle between the incident $X$ and the outgoing $\gamma '$. 
The expected observed event rate as a function of released energy ($E_{rel}\equiv E_X$) is shown in Figure~\ref{fig_2}. 

\subsubsection*{Background evaluation for TAO}

The signals considered in this study are single scattered electrons with energy ranging up to a few MeV. 
The dominant background is the radioactive singles originating from the detector material radio-activities and cosmic-induced particles. 
The radioactive background rate is less than 100 Hz referring to~\cite{tao}. 
Hence we assume the background rate of 100 Hz per ton in order to be conservative.

Due to the high efficiency of veto system on the top of the central detector, the muon induced background is negligible~\cite{tao}. 
In addition, the antineutrino background is also negligible, since the antineutrino inverse beta decay rate from the reactor is about 2000 per day and most of the events have a signature of coincidence. 
Consequently, the only background that has a noticeable impact on our signal measurement is the radioactive singles. 
In order to prevent energy leak, a 5 cm outer sphere is taken as a buffer region resulting in a total fiducial mass of 1.9 tons.

The radioactive background shape was measured in Daya Bay experiment in detail and TAO assumes the same shape~\cite{tao}. 
Both statistical and systematic uncertainties of the radioactive background are considered in this study. 
The systematic uncertainty of the radioactive rate normalization is 
set to 1\%.
This assumption can be obtained from the reactor-off period measured by the same method as the one that is used to measure the radioactive single rate in the reactor-on period. 
According to the TAO analysis group, there is no shape uncertainty assigned to the radioactive background.
The measured signal in TAO is compared to the background uncertainty. 
The excess of measured events over the background uncertainty indicates the existence of hypothetical particles and the corresponding discovery significance can be obtained.
The radioactive background shape below 0.8MeV is not well studied in current reactor neutrino experiments.

\section{Results and discussions}
For the extraction of the confidence level, we studied both the rate-only and the rate+shape measurements. 
Comparing to the signal shape, the background shape across the entire energy range is flat. It is reasonable to imagine that the shape does not contribute significantly to the sensitivity estimate. A separate rate+shape study was preformed and the result is very similar to the rate-only result. Therefore, due to the small impact of the shape information the rate measurement is considered as the main result; i.e. we only use one energy bin for the signal and background.
In parallel, we demonstrate the optimization of the bin width. 
For this purpose the energy range for bin width was changed to provide the highest value of the signal to background ratio.
Since the background shape is flat it is reasonable to choose the bin near the maximum of the detection rate function.
\textcolor{black}{For both cases, the systematic uncertainty on background is set at 1\%.}

A $95 \%$ C.L. for one-sided upper limit can be defined as $1.645\cdot\sqrt{\sigma_{stat}^2+\sigma_{syst}^2}$.
In reactor neutrino experiments, due to the high radioactive
background rate, the systematic uncertainty is dominant in comparison with the statistical one.
Our 100 Hz accidental rate results in a background rate of $1.55\cdot10^9$ over 180 days. 
The systematic uncertainty of the background rate is $\sigma_{syst}=1.55\cdot10^7$, while the statistical uncertainty is $\sigma_{stat}=3.9\cdot10^4$ which is three order of magnitude smaller than $\sigma_{syst}$.
Due to the large systematic uncertainty current LSc reactor neutrino experiments have a limited sensitivity to the listed dark bosons.
A better measurement of the accidental background, which can lead to less systematic uncertainty, can significantly boost the  sensitivity to aforementioned dark bosons.
This critical issue was not mentioned in~\cite{park},  and in our opinion this induces an overestimation of sensitivity in the NEOS experiment.  
Based on the facts mentioned above, the statistical uncertainty posts a negligible impact on the confidence level assessment.

\begin{figure}[ht]
\centering
\includegraphics[scale=0.455]{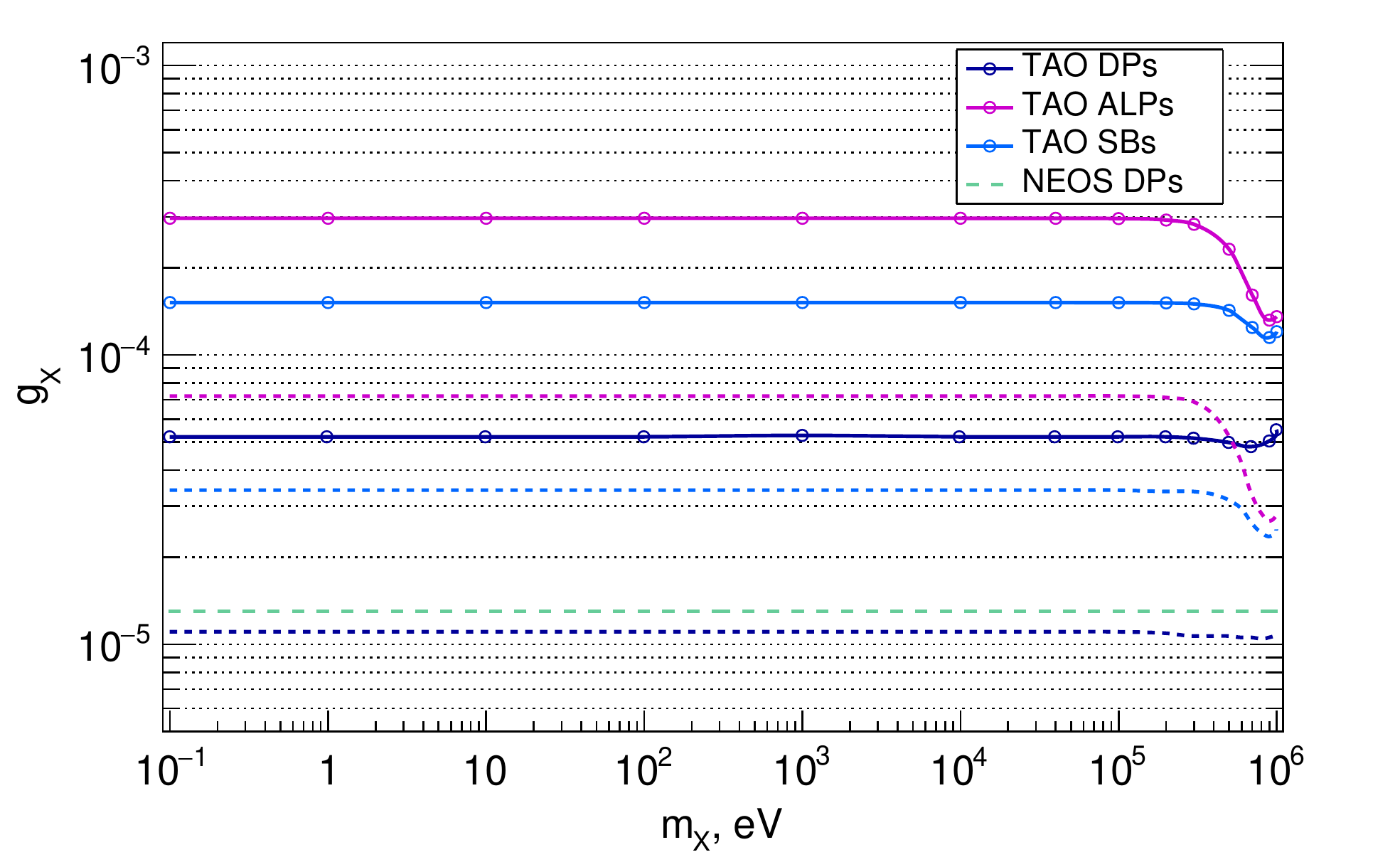}
\caption{ \label{fig_3} Sensitivity to $g_X$ coupling constant at 95\% C.L. as a function of $m_X$ for the TAO detector for DPs, ALPs and SBs. Fiducial volume is 1.9 tons. Exposure time is 180 days for comparison with NEOS~\cite{park} teal dashed line.
The solid lines represent the sensitivity with a single energy bin from 0.8 MeV to 10 MeV.
The dashed lines of the same colors correspond to optimized signal-to-background energy windows for the corresponding boson types.}
\end{figure}

The sensitivity of TAO was calculated for all three types of bosons (DP, ALP, SB) and the result is shown in Figure~\ref{fig_3}.
For the clear comparison between TAO and NEOS the exposure time has been set to 180 fully-duty days.
Even with an enlarged fiducial volume and reactor power, TAO is unable to achieve the same level of sensitivity that was proposed for the NEOS detector because of the systematic uncertainty. It was added in this analysis in comparison with~\cite{park}.
Similar issues were pointed out in~\cite{Danilov:2018bks}.
\begin{figure}[ht]
\centering
\includegraphics[scale=0.44]{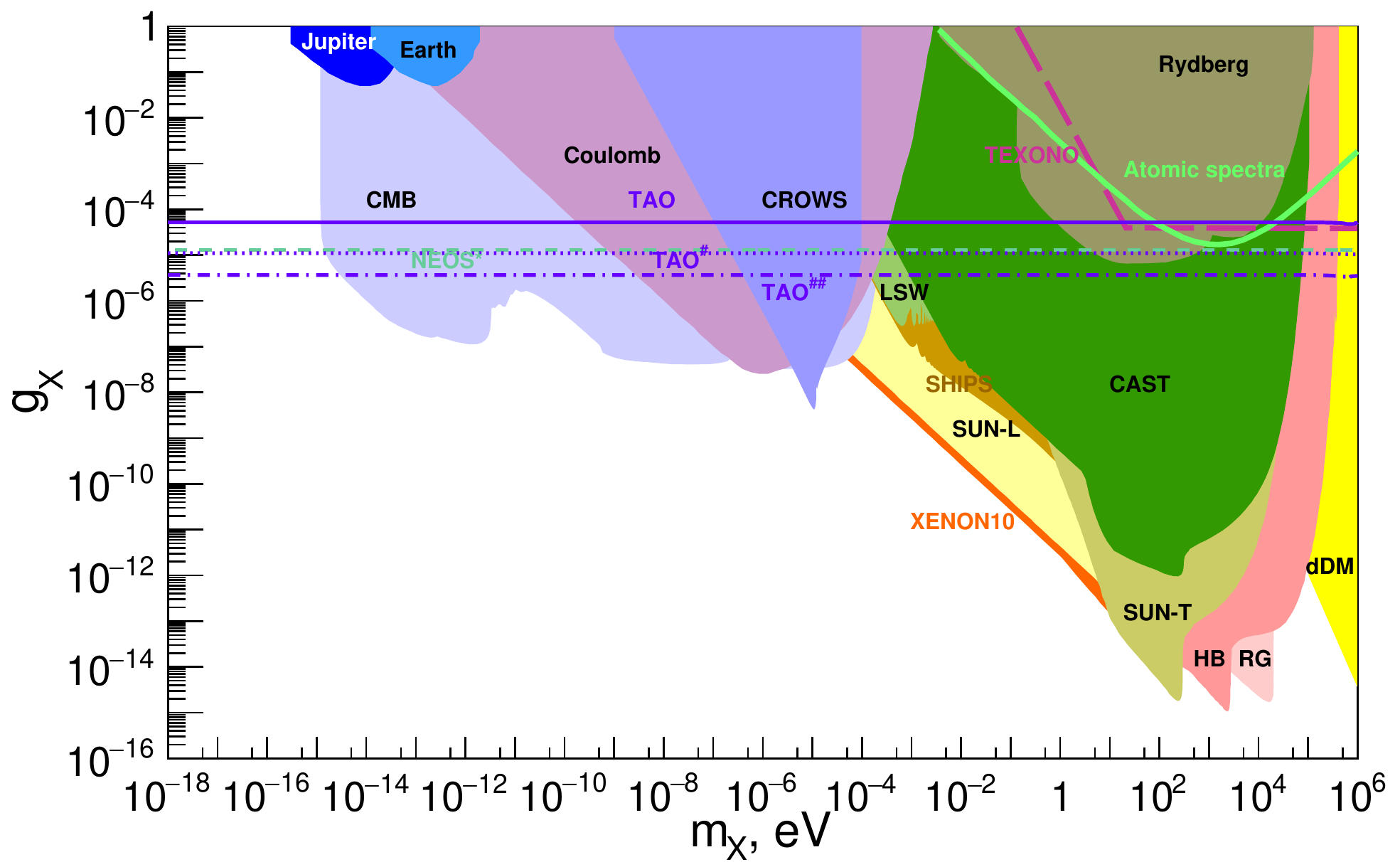}
\caption{ \label{fig_4} Sensitivity comparison at 95\% C.L. of TAO with other experimental results. DP case is only considered. 
 Upper limits are from astrophysics~\cite{Redondo:2013lna}, atomic experiments~\cite{Jaeckel:2010xx}, CAST~\cite{Redondo:2008aa}, CMB~\cite{Mirizzi:2009iz,McDermott:2019lch,Caputo:2020bdy,Garcia:2020qrp}, CROWS~\cite{Betz:2013dza}, decays of relic dark matter~\cite{Redondo:2008ec}, Light Shinning through a Wall (LSW)~\cite{Ehret:2010mh}, SHIPS~\cite{Schwarz:2015lqa}, TEXONO~\cite{Danilov:2018bks} and XENON10~\cite{An:2013yua}.
 The solid line for TAO corresponds to a single energy bin from 0.8 MeV to 10 MeV.
 The single number sign for TAO corresponds to the energy bin range optimized by signal to background ratio.
 The double number sign for TAO corresponds to the energy bin range optimized by signal to background ratio without systematic uncertainty.
 The asterisk for NEOS means the result is controversial.}
\end{figure}
It is interesting to compare our results with other experiments in this field. 
The most detailed results were received in DPs searching, hence we added our sensitivity curve into two dimensional parameter region in Figure~\ref{fig_4}.
As can be seen, all three neutrino experiments NEOS, TAO and TEXONO can not exclude a new area of the parameters space.

As it follows from this research, the searches of light dark bosons with masses below 1 MeV are limited at the level of $\sim10^{-4}$.
The optimization of the signal to background ratio improves the result by an order of magnitude.
Further improvement can not be reached only through increased statistics.
Better knowledge about the radioactive background, especially below 0.8 MeV, and correct optimization of the signal to background ratio are necessary.

\textcolor{black}{For example with} an order of magnitude reduction of the background systematic uncertainty, i.e. a reduction to 0.1\%, the sensitivity of the coupling constant $g_{x}$ increases by a factor of 1.8 for all three hypothetical dark bosons. The sensitivity improvement follows an approximately 4th root of $\alpha$, where $\alpha$ is the background systematic uncertainty decreasing factor.
A significant reduction of the systematic uncertainty is beneficial for such experiments.
Also detailed knowledge about the refueling of the reactor cores will allow us to implement background subtraction. 
And this, in turn, will have a positive effect on sensitivity.
Additionally cross section formulas are provided for further calculations in this area. 

\begin{acknowledgments}

We would like to say great thanks to Dr. Liang Zhan and Dr. Jianrun Hu for providing explicit information and assessment of the JUNO-TAO background.
We also express our special gratitude to Mr. Abraham Teklu for helping us with language editing of the content.
Jiajie Ling acknowledges the support from National Key R\&D program of China under grant NO. 2018YFA0404013, 
and National Natural Science Foundation of China under Grant NO. 11775315.
Jiajun Liao acknowledges the support from the National Natural Science Foundation of China under grant No. 11905299, and Guangdong Basic and Applied Basic Research Foundation under grant No. 2020A1515011479. .
\end{acknowledgments}

\appendix*
\section{Cross sections}
\hypertarget{appendix1}
As it was mentioned in the main text, here we demonstrate complete expressions for the cross sections of all three types of dark bosons.
\subsection*{DPs -- vector bosons}
$A, B, C$ parameters for equation~\eqref{eq_4}:
\begin{widetext}
\begin{equation}
\label{eq_a1}
A=\sqrt{m_e^4-2 m_e^2 (s+m_X^2)+(s-m_X^2)^2}
\cdot\left[m_e^6-m_e^4(s+m_X^2)+m_e^2s(2m_X^2+15s)+s^2(s+7 m_X^2)\right],
\end{equation}
\begin{equation}
\label{eq_a2}
B=-2s^2\left[-3m_e^4+2 m_e^2(m_X^2-3s)+2m_X^4-2m_X^2 s +s^2\right]
\end{equation}
and
\begin{equation}
\label{eq_a3}
C=\frac{m_e^2-\sqrt{m_e^4-2 m_e^2(m_X^2+s)+(s-X^2)^2}-m_X^2+s}{m_e^2+\sqrt{m_e^4-2 m_e^2(m_X^2+s)+(s-X^2)^2}-m_X^2+s}.
\end{equation}
The inverse differential cross section of $X e^-\rightarrow \gamma e^-$ is: 
\begin{eqnarray}
\label{eq_a4}
\frac{d\sigma_{X e^-\rightarrow \gamma e^-}}{dT}&=&\frac{Z\alpha g_X^2}{8 m_e^2 (E_X^2-m_X^2)(E_X-T)^2(2E_Xm_e+m_X^2)^2}
\\\nonumber
&&\times\left[16 E_X^4m_e^3- 16 E_X^2 m_e^3 T(2E_X+m_e)+ 2 m_e m_X^4(3 E_X^2-5 E_X T +m_e^2+2 T (T+m_e))\right.
\\\nonumber
&&\left.+8  m_e^3T^2(3 E_X^2+2 E_Xm_e+m_e^2)+4m_e^2m_X^2(4 E_X^3-8E_X^2 T+E_XT(5T-2m_e)\right.
\\\nonumber
&&\left.+T(2m_e^2+3 m_e T-T^2))+m_X^6(E_X+m_e-T)-8 E_Xm_e^3 T^3\right]\,,
\end{eqnarray}
where $Z$ is the atomic number of LSc (for LAB $Z\approx 3$).
\end{widetext}
 
 \subsection*{ALPs -- pseudoscalar bosons}

The Lagrangian involving pseudoscalar bosons $X$ can be expressed as:
\begin{equation}
\label{eq_a5}
\mathcal{L}\supset  \frac{1}{2}m_X^2 X^2- g_X \bar{e}\gamma^5 e X.
\end{equation}
The differential cross section of $\gamma e^-\rightarrow X e^-$ is:
\begin{equation}
\begin{split}
\label{eq_a6}
&\frac{d\sigma_{\gamma e^-\rightarrow X e^-}}{dx}=\frac{Z\alpha g_X^2}{4 (s-m_e^2)^3(1-x)}\\
&\times\left[x^2(s-m_e^2)^2+2m_X^4-2m_X^2 s x+\frac{2m_e^2m_X^2 x}{1-x}\right].
\end{split}
\end{equation} 
The total cross section of $\gamma e^-\rightarrow X e^-$ is:
\begin{equation}
\label{eq_a7}
\sigma_{\gamma e^-\rightarrow X e^-}=\frac{Z\alpha g_X^2}{8s^2(s-m_e^2)^3}\left[A+B\log C\right],
\end{equation}

\begin{widetext}
where
\begin{equation}
\begin{split}
\label{eq_a8}
A=\sqrt{m_e^4-2 m_e^2 (s+m_X^2)+(s-m_X^2)^2}
\cdot\left[m_e^6-m_e^4(5 s+m_X^2)+m_e^2s(2m_X^2+7s)+s^2(7 m_X^2-3 s)\right],
\end{split}
\end{equation}
\begin{equation}
\label{eq_a9}
B=-2s^2\left[m_e^4-2 m_e^2(m_X^2+s)+2m_X^4-2m_X^2 s +s^2\right]
\end{equation}
and $C$ is the same as~\eqref{eq_a3}.
The inverse differential cross section of $X e^-\rightarrow \gamma e^-$ is: 
\begin{eqnarray}
\label{eq_a10}
\frac{d\sigma_{X e^-\rightarrow \gamma e^-}}{dT}&=&\frac{Z\alpha g_X^2}{16 m_e^2 (E_X^2-m_X^2)(E_X-T)^2(2E_Xm_e+m_X^2)^2}
\\\nonumber
&&\times\left[8 E_X^2 m_e^3 T^2+ 2 m_e m_X^4(E_X^2-E_X T +2 m_e T)-8 E_X m_e^3T^3\right.
\\\nonumber
&&\left.+4m_e^2m_X^2T^2(E_X+m_e-T)+m_X^6(E_X+m_e-T)\right]\,.
\end{eqnarray}
\end{widetext}

 \subsection*{SBs -- scalar bosons}
 The Lagrangian involving scalar bosons $X$ can be expressed as:
\begin{equation}
\label{eq_a11}
\mathcal{L}\supset  \frac{1}{2}m_X^2 X^2- g_X \bar{e} e X.
\end{equation}
\\
\\
The differential cross section of $\gamma e^-\rightarrow X e^-$ is:
\begin{widetext}
\begin{equation}
\label{eq_a12}
\frac{d\sigma_{\gamma e^-\rightarrow X e^-}}{dx}=\frac{Z\alpha g_X^2}{4 (s-m_e^2)^3(1-x)}\Big[m_e^4x^2-2 m_e^2 s(x-4)x+2 m_X^4-2 m_X^2 s x +s^2x^2-\frac{8m_e^4x-2m_e^2m_X^2(5x-4)}{1-x}\Big].
\end{equation} 
\end{widetext}
The total cross section of $\gamma e^-\rightarrow X e^-$ is:
\begin{equation}
\label{eq_a13}
\sigma_{\gamma e^-\rightarrow X e^-}=\frac{Z\alpha g_X^2}{8s^2(s-m_e^2)^3}\left[A+B\log C\right],
\end{equation}

\begin{widetext}
where
\begin{equation}
\begin{split}
\label{eq_a14}
A=\sqrt{m_e^4-2 m_e^2 (s+m_X^2)+(s-m_X^2)^2}
\cdot\left[m_e^6-m_e^4(5 s +m_X^2)+m_e^2s(2m_X^2-25s)+s^2(7 m_X^2-3 s)\right],
\end{split}
\end{equation}
\begin{equation}
\label{eq_a15}
B=-2s^2\left[9 m_e^4+ m_e^2(6 s - 10m_X^2)+2m_X^4-2m_X^2 s +s^2\right]
\end{equation}
and $C$ is the same as~\eqref{eq_a3}.
The inverse differential cross section of $X e^-\rightarrow \gamma e^-$ is: 
\begin{eqnarray}
\label{eq_a16}
\frac{d\sigma_{X e^-\rightarrow \gamma e^-}}{dT}&=&\frac{\alpha g_X^2}{16 m_e^2 (E_X^2-m_X^2)(E_X-T)^2(2E_Xm_e+m_X^2)^2}
\\\nonumber
&&\times\left[8 E_X^2 m_e^3 T(4m_e+T)+ 2 m_e m_X^4(E_X^2-E_X T +2 m_e (T-m_e))-8 E_X m_e^3T^2(4m_e+T)\right.
\\\nonumber
&&\left.- 4m_e^2m_X^2T(-E_X(4m_e+T)+4 m_e^2+3m_e T +T^2)+m_X^6(E_X+m_e-T)-16 m_e^5 T^2\right]\,.
\end{eqnarray}
\end{widetext}

\end{document}